\documentclass{article}

\begin{document}

\title{Quarticles and the Identity of Indiscernibles\footnote{To appear in K.A. Brading and E. Castellani (eds.), "Symmetries in Physics: Philosophical Reflections", CUP 2003.}}

\author{Nick Huggett\\ University of Illinois, Chicago\footnote{Contact the author at huggett@uic.edu}}

\maketitle

\ 

\section{Introduction}In sections 5.3-4 of their paper in this volume\footnote{K.A. Brading and E. Castellani (eds.),   \textit{Symmetries in Physics: Philosophical Reflections}, CUP 2003}, French and Rickles raise the question of the logical relations between the indistinguishability postulate (IP) and the various senses in which particles might fail to be individuals. In section 5.3 they refer to the convincing arguments of French and Redhead (1988) and of Butterfield (1993) that IP does not logically entail non-individuality, understood several ways -- even though, as all seem to concede, there \emph{is} something perverse about taking bosons and fermions to be individuals. Going the other way, as Huggett and Imbo\footnote{N. Huggett and T. Imbo, ``Identicality and indistinguishability'', in preparation.}
 show, if non-individuality is taken to mean the absence of continuous distinguishing trajectories, characteristic of standard quantum mechanics (QM), then non-individuality does not entail IP. Nor, as French and Rickles point out, do substance or haecceity views of individuality.

But what if we conceive of individuality in terms of the principle of the identity of indiscernibles (PII)? First, French and Redhead (1988) and Butterfield (1993) have given theorems showing that bosons and fermions violate PII, while the former have also demonstrated violations of PII in the case of a certain paraparticle state. But these cases, as I will explain (and as French and Rickles point out), cover just a very few of the possible kinds of quantum particles, and so for each kind the question arises as to whether it violates PII. I will give an answer to this question here, rather more general than -- though based on -- those previously offered.

So consider the theorems proven about PII. Suppose one has an $n$-particle system. The states of such a system lie in the Hilbert space that is the tensor product of $n$ Hilbert spaces, one for each of the particles. Suppose further that the Hilbert space for particle $i$ or $j$ -- appearing as the $i^{th}$ and the $j^{th}$ factors of the tensor product space -- is $H$. And suppose finally that $Q$ is an observable on $H$ with eigenvalues $p$ and $q$ (and let $I$ represent the identity on any of the factors). Then, $Q_i = I\otimes I\otimes ... \otimes Q\otimes ... \otimes I$ (with $Q$ as the $i^{th}$ factor) is intuitively the observable corresponding to a measurement of the value of $Q$ for the $i^{th}$ particle, and \textit{mutatis mutandis}  $Q_j$.

Then, French and Redhead (1988) showed the following: for any state $\Psi$ such that $P_{ij}\Psi = \pm\Psi$ (where $P_{ij}$ transposes the $i^{th}$ and the $j^{th}$ factors of the tensor product space) so that $i$ and $j$ are either symmetrized (if $+$) or antisymmetrized (if $-$),\footnote{Note that we have assumed nothing at this stage about whether the particles are identical, or (anti)symmetrized in every state, or what the effect of other permutations of $\Psi$ might be. We will consider these factors below.}

\begin{equation}
\mathrm{pr}^\Psi(Q_i=q) = \mathrm{pr}^\Psi(Q_j=q) \end{equation}

\noindent and

\begin{equation}
\mathrm{pr}^\Psi(Q_i=q|Q_j=p) = \mathrm{pr}^\Psi(Q_j=q|Q_i=p). \end{equation}

\noindent That is, for example, the probability that the system, while in state $\Psi$, would possess value $q$ for observable $Q_i$ if measured is equal to the probability that it would possess that value for $Q_j$ if measured -- roughly, the chance that particle $i$ has value $q$ is the same as the chance that particle $j$ does. Butterfield (1993) extends these results by considering a second observable on $H$, $Q'$, with eigenvalues $q'$ and $p'$ and corresponding observables $Q'_i$ and $Q'_j$ on the tensor product space. He also considers a third particle whose Hilbert space is the $k^{th}$ factor and for which $Q'$ is also an observable.\footnote{Now $k$ need not be identical to the other particles -- the proof certainly does not assume that it is (anti)symmetrized with respect to $i$ or $j$. However, the proof does assume that it has the same Hilbert space $H$, or at the very least that there is a correspondence between the observable $Q'$ on $H$ and an observable on $k$'s Hilbert space.} Then he shows - again for bosons and fermions -- that

\begin{equation}
\mathrm{pr}^\Psi(Q_i=q|Q'_j=p') = \mathrm{pr}^\Psi(Q_j=q|Q'_i=p') \end{equation}

\noindent and

\begin{equation}
\mathrm{pr}^\Psi(Q'_k=q'|Q_i=p) = \mathrm{pr}^\Psi(Q'_k=q'|Q_j=p). \end{equation}

Since the probabilities for the possession of eigenvalues of observables capture the dynamical properties of a quantum system exhaustively, the authors conclude that these four results show that any pair of bosons or fermions in a system are indistinguishable by monadic (the first result) or relational (the other results) dynamical properties. The possibility remains that the particles are distinguished by their intrinsic properties, those that are state-independent, and can be treated as ``c-number'' quantities, such as spin magnitude, mass, charge, or colour. So let's suppose that the particles are ``identical'', sharing all their intrinsic properties. Then, as Butterfield (1993) and French and Redhead (1988) conclude, $i$ and $j$ in state $\Psi$ violate both logically strong -- that individuals sharing all monadic properties are identical -- and weak -- that individuals sharing all monadic \emph{and relational} properties are identical -- forms of the principle of the identity of indiscernibles.

But of course there is more. For if $i$ and $j$ are identical bosons or fermions then $P_{ij}\Psi=\pm\Psi$ is always true: they satisfy the ``symmetrization postulate''. And so, conclude Butterfield and Co., a pair of identical bosons or fermions always violate PII.

Now, a word of clarification is in order. One might well think that if we are, say, talking about bosons, then since the system is only ever in symmetrized states, the appropriate Hilbert space is just the \emph{symmetrized sector} of the tensor product space -- just those states for which $P_{ij}\Psi=\Psi$. But $Q_i$ and $Q_j$ are not observables on the symmetric sector, but only on the full tensor product space, so the proof is not applicable. In fact, $Q_i$ and $Q_j$ don't in general even act as operators on the symmetric sector, since their action on a symmetric vector is to take it out of the sector -- e.g., $Q_i\Psi$ is in general not symmetric. (If they do act as operators then, as it's easy to see, their \emph{restrictions} to the symmetric sector are identical.) Thus one might say that they are not really observables at all for bosons (or at least not different observables). But French and Redhead (1988) are sensitive to this kind of worry, for they argue that one can indeed take the full tensor product space as the state space for the bosons, taking all states, symmetrized or not, to be possible but -- in a terminology I propose -- taking only the symmetric ones to be ``preparable''. After all, they point out, we can understand the unpreparability
 of the unsymmetrized states in terms of a boundary condition (that the initial state was symmetrized) and the symmetrization of the Hamiltonian (which guarantees that the symmetry type is preserved).\footnote{This view has its problems, especially those raised by Redhead and Teller (1991).} So we can -- and will for the purposes of this discussion -- take $Q_i$ and $Q_j$ to be (distinct) operators, even though they may have no restriction to the symmetric sector (or restrictions that are not distinct). Finally, however, one might object that $Q_i$ and $Q_j$ so construed are no longer observables, since they violate the ``indistinguishability postulate'' (IP), according to which every observable must commute with every permutation of identical particles: $[P_{ij},O]=0$ in this case. French and Redhead's response is analogous to their reply to the first objection: they allow ``observables'' that cannot be observed, so that -- in principle -- any Hermitian operator, whether commuting or not, might be used to discern particles. I will follow this suggestion, but (until section 3) I will use ``observable'' to refer \emph{only} to those Hermitian operators that satisfy IP. Hence we want to know whether PII is violated not for the smaller set of properties -- those which correspond to observables -- but for the larger set -- those corresponding to all Hermitian operators -- which contains the smaller as a subset. (Of course if the PII is violated for the latter it is automatically violated for the former, since it is a subset, and hence contains no more properties that might discern particles.) What the results show is that identical bosons are indeed indiscernible (and so are identical fermions).\footnote{Two remarks: First, when we look at other kinds of particles we will find that it is quite possible to consider observables that violate IP, and so we can avoid the debate about whether French and Redhead's scheme is legitimate. Second, Saunders (this volume) argues for a rather different and highly plausible understanding of PII in quantum mechanics (see also French and Rickles, this volume).}

Consider how the proofs of these results go, following the style adopted by Butterfield (1993). What we need in order to calculate these probabilities is the probabilities -- for state $\Psi$ -- of such propositions as the atomic $Q_i=q$ and conjunctive $Q_i=q \ \& \ Q'_j=p$, since the conditional probabilities can be found in terms of these unconditional probabilities. But these quantities can be expressed in terms of the expectation values for the projection operators onto the eigenstates with the appropriate eigenvalues (and products of such operators). For example, let the projection operators onto the subspace of states that are eigenstates of $Q_i $ and $Q'_j$ with eigenvalues $q$ and $p$ respectively be denoted $^qQ_i$ and $^pQ'_j$ respectively. And so on in the obvious way for the other projection operators. Then according to the standard QM algorithm, for instance:

\begin{equation}
\mathrm{pr}^\Psi(Q_i=q) = \langle\Psi|^qQ_i|\Psi\rangle \end{equation}

\noindent and

\begin{equation}
\mathrm{pr}^\Psi(Q_i=q \ \& \ Q'_j=p) = \langle\Psi|^qQ_i\ ^pQ'_j|\Psi\rangle \end{equation}

Finally, the purely algebraic relation $(P_{ij})^2=I$ plus the fact that $Q_j=P_{ij}Q_iP_{ij}$ -- let's call this the `conjugacy condition' (CC) -- which implies that $^qQ_j=P_{ij}\ ^qQ_iP_{ij}$, will deliver our proofs. Given $P_{ij}\Psi=\pm\Psi$,

\begin{eqnarray}
\nonumber\mathrm{pr}^\Psi(Q_i=q) &=& \langle\Psi|^qQ_i|\Psi\rangle\\ \nonumber &=& \langle\Psi|P_{ij}\ ^qQ_jP_{ij}|\Psi\rangle\\ \nonumber &=& \langle\Psi|^qQ_j|\Psi\rangle\\
  &=& \mathrm{pr}^\Psi(Q_j=q)
\end{eqnarray}

\noindent and

\begin{eqnarray}
\nonumber\mathrm{pr}^{\Psi }(Q_{i}=q\ \&\ Q_{j}'=p) &=& \langle \Psi |^{q}Q_{i}\ ^{p}Q_{j}'|\Psi \rangle\\
\nonumber &=& \langle \Psi |P_{ij}\ ^{q}Q_{j}P_{ij}P_{ij}\ ^{p}Q_{i}'P_{ij}|\Psi \rangle\\
\nonumber &=& \langle \Psi |^{q}Q_{j}\ ^{p}Q_{i}'|\Psi \rangle\\ &=& \mathrm{pr}^{\Psi }(Q_{j}=q\ \&\ Q_{i}'=p) \end{eqnarray}

\noindent and so on for any of the atomic or conjunctive propositions that we need in order to derive the conditional probabilities, and hence they are equal too, proving the theorems.

\section{Generalizations}I want to generalize these results still further, in two ways. First, why stop with conditionalizing on only one or two atomic propositions? Why not take the most general relational probability to be

\begin{eqnarray}
\nonumber\mathrm{pr}^{\Psi}(\Pi _{ij})\equiv\mathrm{pr}^{\Psi }(Q_{i}=q_{i}\ \&\ Q_{i}'=p_i\ \&\ ...\ \&\ R_{j}=r_{j}\ \&\ S_{j}=s_{j}\ \&\ ...\\
\nonumber \&\ T_{k}=t_{k}\ \&\ U_{k}=u_{k}\ \&\ ...\ |V_{i}=v_{i}\ \&\ W_{i}=w_{i}\ \&\ ...\\
X_{j}=x_{j}\ \&\ Y_{j}=y_{j}\ \&\ ...\ \&\ Z_{k}=z_{k}\ \&\ A_{k}=a_{k}\ \&\ ...\ )
\end{eqnarray}

\noindent where all the operators are of the same form as $Q_i$, with a suitable Hermitian operator on $H$ replacing $Q$ in the appropriate ``slot'' in the tensor product.\footnote{Note that in this case, and as we go on, the observables in question -- and hence their projection operators -- need not all commute. Thus the conditional probability, found as before in terms of the expectation values of the corresponding products of projection operators is dependent on the order of the operators. We will adopt the convention that the order of the projection operators is the same as the order of the propositions in the expression for the probability.} (And why not let the ellipses be filled in with propositions concerning a fourth and fifth and so on particles?) Then we need to show that if $P_{ij}\Psi=\pm\Psi$ then $\mathrm{pr}^\Psi(\Pi_{ij}) = \mathrm{pr}^\Psi(\Pi_{ji})$, where the latter argument is obtained from $\Pi_{ij}$ by transposing the $i$ and $j$ labels on the observables throughout. And to show this we need -- much as in the cases considered so far -- to show that the probability of an arbitrary conjunction of atomic propositions is equal to that of its transposition. And to do this we need one more fact about these operators, namely that for $k\neq i,j$, $P_{ij}Q_kP_{ij}=Q_k$, from which it follows that $P_{ij}\ ^qQ_kP_{ij}=\ ^qQ_k$ or $P_{ij}\ ^qQ_k=\ ^qQ_kP_{ij}$ (and similarly for any of the other operators in question) -- let's call this the ``independence condition' (IC).\footnote{In fact, as you will have found if you've worked through the proofs, this condition is also required to prove the third and fifth results of Butterfield and Co.} Then we see for example that

\begin{eqnarray}
\label{blah1}
\nonumber\mathrm{pr}^{\Psi }(Q_{i}=q_{i}\ \&\ Q_{i}'=p_i\ \&\ ...\ \&\ R_{j}=r_{j}\ \&\ S_{j}=s_{j}\ \&\ ...\&\ T_{k}=t_{k}\ \&\\ \nonumber U_{k}=u_{k}\ \&\ ...\ ) \\
=\langle \Psi |^{q_{i}}Q_{i}\ ^{p_i}Q_{i}'\ ...\ ^{r_{j}}R_{j}\ ^{s_{j}}S_{j}...^{t_{k}}T_{k}\ ^{u_{k}}U_{k}\ ...\ |\Psi \rangle  \\ =\langle \Psi |P_{ij}\ ^{q_{i}}Q_{j}P_{ij}P_{ij}\ ^{p_i}Q_{j}P_{ij}\ ...\ P_{ij}\ ^{r_{j}}R_{i}P_{ij}P_{ij}\ ^{s_{j}}S_{i}P_{ij}\ ...\ ^{t_{k}}T_{k}\ ^{u_{k}}U_{k}\ ...\ |\Psi \rangle \end{eqnarray}

\noindent by CC

\begin{equation}
= \langle\Psi|P_{ij}\ ^{q_i}Q_j\ ^{q'_i }Q'_j\ ...\ ^{r_j}R_i\ ^{s_j}S_i\ ...\ ^{t_k}T_k\ ^{u_k}U_k\ ...\ P_{ij}|\Psi\rangle, \end{equation}

\noindent using $P_{ij}^2=I$ and applying IC to shift the remaining $P_{ij}$ all the way to the right, to obtain for (anti)symmetrized $\Psi$

\begin{eqnarray}
=\langle \Psi |^{q_{i}}Q_{j}\ ^{p_i}Q_{j}'\ ...\ ^{r_{j}}R_{i}\ ^{s_{j}}S_{i}\ ...\ ^{t_{k}}T_{k}\ ^{u_{k}}U_{k}\ ...\ |\Psi \rangle \\
\nonumber =\mathrm{pr}^{\Psi }(Q_{j}=q_{i}\ \&\ Q_{j}'=p_i\ \&\ ...\ \&\ R_{i}=r_{j}\ \&\ S_{i}=s_{j}\ \&\ ... \&\ T_{k}=t_{k}\ \&\\
\label{blah2} U_{k}=u_{k}\ \&\ ...\ ),
\end{eqnarray}

\noindent as we required. So the upshot is that no conditional probabilities for the system taking on values for the Hermitian operators in question will serve to discern identical bosons or fermions in a system.

There is, however, a further generalization that I think should be made, one which also preserves the result regarding PII. Why assume that only operators of the form $I\otimes ... \otimes Q\otimes ... \otimes I$ can represent single particle properties? Of course they do, but, I would suggest, as a special case. Here's the more general notion that I have in mind. Suppose that for an $n$-particle system we find a ``family'' of $n$ Hermitian operators, $\{O_1,...O_i,...O_j...O_n\}$, that satisfy CC pair-wise: $O_j=P_{ij}O_iP_{ij}$ for all $i$, $j$. Then it seems to me that they are candidates for representing single particle properties, since the natural interpretation is that whatever quantity $O_i$ represents of the $i^{th}$ particle, $O_j$ represents of the $j^{th}$ particle, and so on. Clearly $Q_i$, $Q_j$ and so on satisfy this condition, but it allows other possibilities too: for example, $A\otimes A\otimes ... \otimes B\otimes ... \otimes A$ ($A$, $B$ observables on $H$) and certain sums of such operators. Now one might argue that being a \emph{single particle} Hermitian operator requires more than membership in such a family (that all or some members of the family are distinct, for example) but no matter for our purposes: the ``minimal condition'' of \emph{membership in such a family of single particle operators} is enough to prove all our results. For these results only depend on two properties of the single particle operators: CC and IC. CC for $Q_i$ and $Q_j$ etc. follows immediately (as a special case) of the minimal condition, and IC can be easily shown. Suppose that the minimal condition is satisfied by Hermitian operators $\{O_1, ... ,O_n\}$. Then

\begin{equation}
O_k = P_{jk}O_jP_{jk} = P_{ij}P_{ik}P_{ij}O_jP_{ij}P_{ik}P_{ij} = P_{ij}P_{ik}O_iP_{ik}P_{ij} = P_{ij}O_kP_{ij}, \end{equation}

\noindent where the second step uses the algebraic identity $P_{jk} = P_{ij}P_{ik}P_{ij}$, which holds if $k \neq i,j$, and the other steps by the minimal condition. And so the general proof goes through -- for bosonic and fermionic $i$, $j$ -- as long as the minimal condition is satisfied; so identical bosons and fermions are indistinguishable by anything we might consider to be a single particle property.

This way of constructing the proofs is nice, I think, because it shows very clearly just what assumptions are doing the work, and because it points the way to extending these considerations to kinds of particles other than bosons and fermions, as French and Rickles suggest.

\section{Quarticles}\label{quarticles}First though, what are these other kinds of particles? As French and Rickles describe (and see for example Greiner and Muller, 1989, chapter 9, for more details), the bosonic and fermionic representations of the permutation group $S_n$ are only two possible representations -- the 1-dimensional ones. As the number of particles increases, one finds representations of higher and higher dimensions. This raises the question of whether there could be a species of quantum particle for every representation -- as bosons and fermions correspond to the symmetric and antisymmetric representations -- and indeed whether there could be a species of particle for every direct sum of representations. This is the question explored and answered by Hartle, Stolt and Taylor (e.g. 1970). More precisely, we want to characterize each (statistical) kind of particle by a rule that specifies, for each $n$, what the space of allowed states is -- after all, a species of particle can come in any number. In this case it is natural to impose a condition of ``cluster decomposition''. First, suppose that one looks at $m$ of the particles in a system of $n$: the rule for $n$ such particles determines their state space, from which we can extract the state space of the first $m$, which must carry exactly the same representations as the rule for $m$ particles demands. Second, in the other direction, if we take two systems of $m$ and $n$ particles respectively, then the rule determines their state spaces, and then the joint state space obtained from them must satisfy the rule for $n+m$ particles. That is, cluster decomposition demands that parts and wholes must be related consistently by the state space rule for any species of particle.

Hartle, Stolt and Taylor showed that this condition entails a very simple classification of the species of possible particles. They found a correspondence between pairs of natural numbers $(p,q)$ satisfying $p+q>0$  \emph{together with} $\infty$ and the allowed species of particles. While these numbers determine which representations are allowed to any species of particle in a fairly direct way, space does not permit that I explain how here.\footnote{See for example N. Huggett, in preparation, ``What is an elementary quarticle?'', appendix, for an account.} There are, however, a few points and illustrations that I can usefully make.

First, for bosons (or fermions) all Hermitian operators on the space of symmetric (or antisymmetric) states satisfy IP, so all are observables as we defined them earlier. In the higher dimensional representations this is not so: there are Hermitian operators on a $(p,q)$ state space that do not commute with all permutations. (Though that the energy be one of these operators is sufficient to guarantee again that the system will never evolve out of the allowed state space.) Hence we should no longer define ``observables'' to be only those operators that satisfy IP -- any Hermitian operator on the allowed space could in principle represent a distinct observable quantity. And so corresponding to (almost) every $(p,q)$ and $\infty$ there is both a possible ``distinguishable'' kind of particle for which all Hermitian operators on the space are observables and a distinct possible ``indistinguishable'' kind of particle for which IP is imposed on observables.\footnote{R. Espinoza, T. D. Imbo and M. Satriawan, ``Identicality, (in)dis\-ting\-uish\-ability and quantum statistics'', in preparation, ask what other sets of Hermitian operators -- fewer than all those on the state space but more than just those that satisfy IP -- can be taken as the observables, and show how to classify all possible answers. Huggett and Imbo, `Identicality and indistinguishability', in preparation, demonstrate why quantum mechanics in no way entails IP.} That is, in choosing the properties of a type of particle, nature needs to decide not just $p$ and $q$, but also whether any physical quantities correspond to observables violating IP.

Then, for $\infty$ particles the whole tensor product space is the state space, and if they are distinguishable then they are called ``quantum Maxwell-Boltzmann particles'', since the number of independent states available to them is always the same as the number of corresponding classical states. Bosons are $(1,0)$ and fermions $(0,1)$ particles, and they are the only cases -- because all operators on the (anti)symmetric sector satisfy IP -- for which there is no distinguishable kind. ``Parabose'' particles of order $p$ (those for which up to $p$ particles may be mutually antisymmetrized) are $(p,0)$ particles, while ``parafermi'' particles of order $q $ are $(0,q)$ particles. Now it is usual to speak of these last two kinds of particles as ``paraparticles'' but we are without a convenient terminology to refer to all particles in the $(p,q)$ classification (other than the bland ``$(p,q)$-particles''). To show that these kinds of particles interpolate between quanta (bosons and fermions) and classical(-like) particles (quantum Maxwell-Boltzmann particles) let's call them ``quarticles'', of which paraparticles are a special class. And so, French and Rickles's question concerns PII for quarticles: When (if ever) does it hold? When (if ever) does it fail?\footnote{French and Redhead (1988) consider a specific 3-particle state in a 2-dimensional representation (which is allowed to infinitely many kinds of quarticle) and show that two of the particles are indistinguishable while either can be distinguished from the third.}
``
Here's a (fairly) comprehensive answer to this question, in three parts:

\

\noindent (i) For any number of any kind of identical quarticle (even as a subsystem of a system including other non-identical particles) there are states in which no particles are discernible. This follows because for any quarticles either totally symmetrized or totally antisymmetrized states (for which $P_{ij}\Psi=\pm\Psi$, where $i$, $j$ label \emph{any} pair of the identical quarticles) are allowed, and we have already seen that they are states of indiscernible particles.

\

\noindent (ii) For any number greater than two of any kind of identical quarticle but quanta there are states in which some particles are discernible and some are indiscernible.

\

Suppose there are $m$ identical quarticles. Again let $Q$ be a Hermitian operator on the single particle Hilbert space $H$, and define  $Q_1$ etc. as before (i.e. $Q_1=Q\otimes I\otimes...\otimes I$ etc.), but now supposing that $Q$ is non-degenerate with eigenstates $\phi_1$, $\phi_2$, ..., $\phi_m$, ... with corresponding eigenvalues $q_1$, $q_2$, ..., $q_m$, ... (so that $H$ is at least $m$-dimensional). Let $S_{ab...c}$ be the operator on the tensor product space that totally symmetrizes states with respect to transpositions of the $a^{th}, b^{th},... c^{th}$ factors, and similarly let $A_{ab...c}$ be the operator on the tensor product space that totally \emph{anti}symmetrizes states with respect to transpositions of the $a^{th}, b^{th},... c^{th}$ factors.\footnote{Thus, for example, $S_{12...m}\phi_1\phi_2...\phi_m...\phi_n$ is the normalized sum of every permutation of the first $m$ factors of $\phi_1\phi_2...\phi_m...\phi_n$, and
$A_{12...m}\phi_1\phi_2...\phi_m...\phi_n$ the normalized sum of the even permutations minus the odd permutations of the first $m$ factors. For further discussion and the explicit form of these operators see for example Greiner and Muller (1989), chapter 9.}

Then for any system including $m>2$ quarticles of any kind but bosons and fermions either

\begin{equation}
\label{SA}
\Psi_s = S_{23...m}A_{12}\phi_1\phi_2...\phi_m
  \end{equation}

\noindent or

\begin{equation}
\Psi_a = A_{23...m}S_{12}\phi_1\phi_2...\phi_m
  \end{equation}

\noindent is an allowed state (of course \emph{not} the only kind of allowed state). In either case the total (anti)symmetrization of quarticles $2, 3, ..., m$ entails that for $2\leq i,j\leq m$, $P_{ij}\Psi=\pm\Psi$; so from our earlier proofs quarticles $2, 3, ..., m$ are indiscernible. However, we can easily calculate $\mathrm{pr}^\Psi(Q_1=q_1)$ and $\mathrm{pr}^\Psi(Q_i=q_1)$ to see that these probabilities discern the quarticle $1$ from every other quarticle. For example,

\begin{equation}
\Psi_s=S_{2...m}(\phi_1\phi_2...\phi_m - \phi_2\phi_1...\phi_m), \end{equation}

\noindent a sum of $2\cdot(m-1)!$ terms with equal coefficients (one for each permutation of the factors $2,3,...m$). $(m-1)!$ of the terms have $\phi_1$ as the first factor -- all those with positive coefficients. But only $(m-2)!$ of the terms have $\phi_1$ as the $i^{th}$ factor -- the number of permutations of the term with the negative coefficient which have $\phi_1$ in the $i^{th}$ place. Therefore,

\begin{equation}
\mathrm{pr}^{\Psi_s}(Q_1={q_1}) = \langle\Psi|^{q_1}Q_1|\Psi\rangle = (m-1)!/(2\cdot(m-1)!)=1/2
\end{equation}

\noindent while (for $2\leq i\leq m$)

\begin{equation}
\mathrm{pr}^{\Psi_s}(Q_i=q_1) = \langle\Psi|^{q_1}Q_1|\Psi\rangle = (m-2)!/(2\cdot(m-1)!)=1/(2\cdot(m-1)),
\end{equation}

\noindent which are never equal for $m>2$. Hence quarticles 1 and 2 are indeed discernible (as they are in state $\Psi_a$).

\

\noindent (iii) For any number greater than two of any kind of identical quarticle but quanta there are states in which all the particles are discernible.

\

The following is an allowed state for $m>2$ of any kind of quarticle but bosons and fermions:

\begin{equation}
\Psi_d = \sum_{i=1}^m
S_{12...(i-1)(i+1)...m}A_{i(i+1)}\phi_1\phi_2...\phi_m, \end{equation}

\

\noindent where $i\pm1$ is taken modulo $m$. That is, $\Psi_d$ is a sum of states similar to those considered in equation (\ref{SA}); for each of the first $m$ factors of the tensor product, there is contribution to $\Psi_d$ in which the factor and its successor are antisymmetrized before the remaining factors are symmetrized. Then -- as in the proof of (ii) -- we count the total number of terms in the sum in which $\phi_i$ appears as the $i^{th}$ (or $j^{th}$) factor to find the numerator for $\mathrm{pr}^{\Psi_d}(Q_i=q_i)$ (or
$\mathrm{pr}^{\Psi_d}(Q_j=q_i)$) and -- since all the coefficients are equal -- the denominator is just the total number of terms in the sum. Omitting the details we find that:

\begin{equation}
\mathrm{pr}^{\Psi_d}(Q_i=q_i) = \{(m-1)! + (2m-3)\cdot(m-2)!\}/(2\cdot m!) \end{equation}

\noindent while for $i\neq j$

\begin{equation}
\mathrm{pr}^{\Psi_d}(Q_j=q_i) = \{(2m-4)\cdot(m-2)!\}/(2\cdot m!), \end{equation}

\noindent which are unequal for all $m$. Hence every quarticle is discernible from every other by the probability for possessing an appropriate quantity.

\section{Concluding remarks}In conclusion there are a couple of remarks I'd like to add. First, I have addressed the question of whether quarticles are discernible by any Hermitian operators, since this is how the issue was originally raised. But what if we consider instead just those Hermitian operators that satisfy IP, as the $Q_i$ of the proofs do not? To indicate that we are now asking a different question we should have a new name for the principle at stake. The principle is that for every pair of (identical) quarticles there is some conditional probability for the possession of single particle \emph{observable} quantities on which they disagree: let's call this the ``principle of the identity of \emph{indistinguishables}''. (The question of whether this principle holds for a particular system depends on what the observables are for that system; I've assumed that any Hermitian observable that satisfies the IP is an observable, but other choices can be investigated.) We will investigate whether this principle holds exactly as before, except we demand that the probabilities we consider are only for observable quantities.

If the IP is imposed on observables then we will find that the identity of indistinguishables is maximally violated -- any pair of identical quarticles in a system will have all probabilities for the possession of observable quantities in common (just like bosons and fermions).\footnote{This is easy to see intuitively: if the minimal condition for being a single particle observable -- inclusion in a family of observables related by CC pair-wise -- is satisfied, then we have for all $i$, $j$, $O_j=P_{ij}O_iP_{ij}$. But by IP $P_{ij}O_iP_{ij}=O_i$ so $O_i=O_j$. Thus the probability for the $i^{th}$ quarticle to possess some given value of whatever quantity $O_i$ represents must be the same as the probability for the system to possess the same value for $O_j$, which represents the same property but for the $j^{th}$ particle. For the conjunctive probabilities required to find the more general probabilities note that if a Hermitian operator satisfies the IP so do the corresponding projection operators and products of projection operators. Thus a calculation like that of equations \ref{blah1}-\ref{blah2} shows that IP is sufficient to show that for any $\Psi$, $\mathrm{pr}^\Psi(\Pi_{ij})=\mathrm{pr}^\Psi(\Pi_{ji})$ as required.}

The second remark is a question. Our calculations made clear that the (anti)symmetrization of a pair of quarticles is sufficient for a violation of PII. But what about the converse? If none of the conditional probabilities for the possession of values of any Hermitian operators will discern particles $i$ and $j$ in some state, $\Psi$, can we infer that $P_{ij}\Psi=\pm\psi$? Consider, for example, a two particle case in which $Q_1=Q\otimes I$ and $Q_2=I\otimes Q$, with $\phi_i$ the eigenstate of $Q$ with eigenvalue $q_i$. Then if $\Psi=\phi_1\phi_2+e^{i\theta}\phi_2\phi_1$ it is easy to check that, for instance,
$\mathrm{pr}^\Psi(Q_1=q_1)=\mathrm{pr}^\Psi(Q_2=q_1)$. And so even though $P_{ij}\Psi\neq\pm\Psi$ for $\theta\neq n\pi/2$, the Hermitian operators we used to discern quarticles will not discern these two particles. So the question is, even though these probabilities do not discern the particles, does $P_{ij}\Psi\neq\pm\Psi$ entail that \emph{some} conditional probability for the possession of values for some single particle Hermitian operators will discern them? If so then we have the most complete account of PII in many particle quantum mechanics: a pair of identical particles is indiscernible just in case they are mutually (anti)symmetrized.

\section{Appendix}

Since this paper first appeared, I have been able to show that the answer to the question posed in the final paragraph is `yes': if a pair of identical particles is indiscernible then they are mutually (anti)symmetrized.

\ 

\noindent\textbf{Proof:} Suppose that particles $i$ and $j$ are indiscernible in $n$-particle state $\Psi$, so that they are not discerned by appropriate conditional probabilities; then in particular for $n$ observables $Q, Q', \dots R,\dots S, \dots T$ on $H$ with eigenvalues $q, q', \dots r,\dots s, \dots t$ respectively:

\begin{eqnarray}
\label{new1}
\mathrm{pr}^\Psi(Q_1=q\ \&\ Q'_2=q'\ \& \dots R_i=r\ \& \dots S_j=s\ \& \dots T_n=t) \\
= \mathrm{pr}^\Psi(Q_1=q \&\ Q'_2=q'\ \& \dots R_i=r\ \& \dots S_j=s\ \& \dots T_n=t),
\end{eqnarray}

\noindent where is the $n$-fold tensor product $Q_1 = Q\otimes I\otimes I\otimes\dots \otimes I$, and so on for the other observables. But these probabilities are, as usual, given by the expectation values for appropriate products of projection operators; hence (\ref{new1}) can be rewritten as

\begin{eqnarray}
\label{new2}
\langle\Psi | ^{q}Q_1 \ ^{q'}Q_2' \dots ^{r}R_i \dots ^{s}S_j \dots ^{t}T_n|\Psi\rangle \\
= \langle\Psi |^{q}Q_1 \ ^{q'}Q_2' \dots ^{r}R_j \dots ^{s}S_i \dots ^{t}T_n|\Psi\rangle \\
= \langle P_{ij}\Psi | ^{q}Q_1 \ ^{q'}Q_2' \dots ^{r}R_j \dots ^{s}S_i \dots ^{t}T_n|P_{ij}\Psi\rangle,
\end{eqnarray}

\noindent where the last step follows from simple algebra of $P_{ij}$ involving CC and IC. However, operators of the form $^qQ_1 \ ^{q'}Q_2' \dots ^rR_i \dots ^sS_j \dots ^tT_n$ form a basis for $Herm(H_n)$, the Hermitian operators on $H_n$.\footnote{$Herm(H_n)$ contains $n$-fold tensor products of Hermitian operators on $H$ and their sums over the reals; since $^qQ_1$ has the form $^qQ\otimes I\otimes I\otimes\dots\otimes I$ the operator products in (\ref{new2}) have the form $^qQ_1 \otimes ^{q'}Q_2' \otimes\dots\otimes ^rR_i \otimes\dots\otimes ^sS_j \otimes \dots \otimes ^tT_n$, linear combinations of which form any $n$-fold tensor products of Hermitian operators on $H$.} Thus, since any $A\in Herm(H)$ can be written as a linear combination of the operator products in equation (\ref{new2}), for any $A$

\begin{equation}
\langle\Psi |A|\Psi\rangle = \langle P_{ij}\Psi |A|P_{ij}\Psi\rangle.
\end{equation}

\noindent But any states lying in distinct rays differ in the expectation values of some Hermitian operator (for instance, the projection operator onto one of the rays), hence $P_{ij}\Psi = \lambda\Psi$. But the eigenvalues of $P_{ij}$ are $\pm1$, so. QED.

To reiterate then, two identical particles, $i$ and $j$ are indiscernible in state $\Psi$ if and only if $\Psi = \pm P_{ij}\Psi$.

\

\begin{center}

{\bf {\Large References}}
\end{center}

Butterfield, J. (1993). Review article: ``Interpretation and identity in quantum theory''. \textit{Studies in the History and Philosophy of Science},  \textbf{24}, 443-76.

\

French, S., and Redhead, M. (1988). ``Quantum physics and the identity of indiscernibles''. \textit{British Journal for the Philosophy of Science},  \textbf{39}, 233-46.

\

Greiner, W., and Muller, B. (1989). \textit{Quantum mechanics: symmetries}. New York: Springer-Verlag.

\

Hartle, J. B., Stolt, R. H., and Taylor, J. R. (1970). \textit{Phys. Rev. D}, \textbf{2}, 1759.

\

Redhead, M., and Teller, P. (1991). ``Particle labels and the theory of indistinguishable particles in quantum mechanics''. \textit{British Journal for the Philosophy of Science}, \textbf{43}, 201-18.

\end{document}